\documentclass[12pt,preprint]{aastex}

\slugcomment{accepted for publication in AJ}

\shorttitle{SMA observation of CRL 618}
\shortauthors{Nakashima et al.}

\begin{document}


\title{Submillimeter Array Observation of the Proto-Planetary Nebula CRL~618 in the CO $J=6$$-$$5$ Line}


\author{Jun-ichi Nakashima\altaffilmark{1}, David Fong\altaffilmark{2}, Tatsuhiko Hasegawa\altaffilmark{1}, Naomi Hirano\altaffilmark{1},\\ Nico Koning\altaffilmark{3}, Sun Kwok\altaffilmark{1,4}, Jeremy Lim\altaffilmark{1}, Dinh-Van-Trung\altaffilmark{1} and Ken Young\altaffilmark{5}}

\altaffiltext{1}{Academia Sinica Institute of Astronomy and Astrophysics, P.O. Box 23-141, Taipei 106, Taiwan; email(JN): junichi@asiaa.sinica.edu.tw}

\altaffiltext{2}{Harvard-Smithsonian Center for Astrophysics, Submillimeter Array, 645 North A'ohoku Place, Hilo, HI 96720}

\altaffiltext{3}{Department of Physics and Astronomy, University of Calgary, Calgary, Canada T2N 1N4}

\altaffiltext{4}{University of Hong Kong, Faculty of Science, Chong Yuet Ming Physics Building, Hong Kong, China, PR}

\altaffiltext{5}{Harvard-Smithsonian Center for Astrophysics, 60 Garden Street, Cambridge, MA 02138}


\begin{abstract}
We report on the results of a Submillimeter Array interferometric observation of the proto-planetary nebula CRL~618 in the $^{12}$CO $J=6$$-$$5$ line. With the new capability of SMA enabling us to use two receivers at a time, we also observed simultaneously in the $^{12}$CO $J=2$$-$$1$ and $^{13}$CO $J=2$$-$$1$ lines. The $^{12}$CO $J=6$$-$$5$ and $^{13}$CO $J=2$$-$$1$ lines were first interferometrically observed toward CRL~618. The flux of the high velocity component of the $^{12}$CO $J=6$$-$$5$ line is almost fully recovered, while roughly 80\% of the flux of the low velocity component is resolved out. The low recovery rate suggests that the emission region of the low velocity component of the $^{12}$CO $J=6$$-$$5$ line is largely extended. Continuum emission is detected both at 230 and 690 GHz. The flux of the 690~GHz continuum emission seems to be partially resolved out, suggesting dust emission partly contaminates the 690~GHz continuum flux. The cavity structure, which has been confirmed in a previous observation in the $^{12}$CO $J=2$$-$$1$ line, is not clearly detected in the $^{12}$CO $J=6$$-$$5$ line, and only the south wall of the cavity is detected. This result suggests that the physical condition of the molecular envelope of CRL~618 is not exactly axial symmetric. 
\end{abstract}


\keywords{stars: AGB and post-AGB ---
stars: carbon ---
stars: imaging ---
stars: individual (CRL618) ---
stars: kinematics ---
stars: winds, outflows}


\section{Introduction}
Stellar evolution from the asymptotic giant branch (AGB) phase to the planetary nebula (PN) phase is very rapid. This transient phase between the AGB and PN phases is often called the proto-planetary nebula (PPN) phase and/or post-AGB phase \citep{kwo93,van03}. PPNe are considered to play an important role in a wide variety of astrophysical problems: for example, the shape and shaping of PNe, chemical evolution of the evolved stars, and synthesis of organic matter in space \citep{bal02,kwo04}. Unfortunately, however, the number of PPNe identified is rather limited mainly due to their transient time \citep[see, e.g.,][]{szc07}. 

In the last few decades, the well-known example of a PPN, CRL~618 ($=$~RAFGL~618 $=$~IRAS~04395+3601 $=$~Westbrook Nebula), has provided a unique opportunity to investigate the nature of at least one particular case of PPNe. CRL~618 entered its PPN phase about 100 years ago \citep{kwo84}, and the central B0 star is surrounded by a compact HI\hspace{-.1em}I region (its angular size is roughly 0.2$''$$\times$0.4$''$) visible through centimeter- and millimeter-wave continuum emission \citep{wyn77,kwo81,kwo84,mar93}. The flux of the free-free emission at $\lambda=1.3$ and 3~mm has been continuously changing over the last 25 years \citep[see, e.g.,][]{san04}, implying the rapid evolution of this object. The envelope of CRL~618 includes a rich set of molecular species \citep[e.g.,][]{buj88,cer89,fuk94,cer01,cer02,rem05}, and has been repeatedly observed in molecular radio lines, especially that of CO \citep[e.g.,][]{shi93,yam94,haj96,san04}. 

Through radio observations in molecular rotational lines, four different kinematical components have been identified in the molecular envelope of CRL~618. In Figure 1, we present a schematic view of the molecular envelope of CRL~618. The first component (corresponding to (1) in Figure 1) exhibits an almost round shape in CO maps, and its angular size is larger than 20$''$. The expanding velocity of this spherical component is roughly 17.5 km~s$^{-1}$, and is interpreted as the remnant of the spherical mass-loss process of the central star during the AGB phase. The second component (corresponding to (2) in Figure 1) exhibits a bipolar flow observed up to axial distances of $\pm2.5''$ from the nebula center. The expansion velocity of this component reaches up to 350 km~s$^{-1}$. Incidentally, the proper motion of this high velocity bipolar flow has been measured by \citet{san04a}, and the value is $\sim 0.045''$ yr$^{-1}$ corresponding to a distance of 900 pc if we assume the inclination angle of the polar axis is 30$^{\circ}$; this distance is consistent with that estimated by an independent method \citep{goo91}. An additional two kinematical components were recently found by \citet{san04} in their high-resolution ($\sim1''$) interferometric radio observation: an extended structure elongated in the polar direction up to an axial distance of $\pm6''$ from the nebula center, which is expanding at a velocity of about 22 km~s$^{-1}$ (corresponding to (3) in Figure 1) and a dense, inner torus-like core expanding at a velocity less than 12 km~s$^{-1}$ (corresponding to (4) in Figure 1).

Although numerous observational and theoretical efforts have been made on CRL~618, several interesting puzzles remain. First, the largely extended torus-like structure is detectable in the HCO$^{+}$ $J=1$$-$$0$ line, but not in the CO $J=2$$-$$1$ line \citep{san04,san04a}. \citet{san04} suggested that a passage of a shock across the torus-like structure plays an important role in explaining this phenomenon. Second, the light curve of the intensity of the free-free emission exhibits a sudden increase over the past 5 years \citep{san04}, implying that CRL~618 exhibits the activity of a post-AGB wind, which is very poorly understood \citep[e.g.,][]{buj94,jur00,ste01,gar05}. Third, the molecular envelope of CRL~618 exhibits a hybrid chemistry simultaneously including both oxygen- and carbon-rich environments \citep[e.g.,][]{buj88,cer89}. Such a mixture of the C/O-rich chemistry has been recognized in some late-type stars \citep[e.g., CRL~2688, OH~231.8+4.2 and IRAS~19312+1950, ][]{luc86,mor87,nak00,nak04a,nak05,mur07}, but the origin of the mixture is still unclear. 

Further observations of CRL~618, especially using new instruments with new capabilities, would be valuable in understanding the nature of this interesting object. In this paper, we report the first results of a Submillimeter Array\footnote{The Submillimeter Array is a joint project between the Smithsonian Astrophysical Observatory and the Academia Sinica Institute of Astronomy and Astrophysics, and is funded by the Smithsonian Institution and the Academia Sinica.} (SMA) interferometric observation of CRL~618 in the $^{12}$CO $J=6$$-$$5$ line using the new capability of SMA enabling us to observe at 690 GHz. This observation has been made as a part of the SMA 690 GHz observation campaign. In addition to the $^{12}$CO $J=6$$-$$5$ line, we briefly report the results of observations in other molecular lines including the $^{12}$CO $J=2$$-$$1$ and $^{13}$CO $J=2$$-$$1$ lines. The outline of the paper is as follows: in Sect 2, details of the observations and data reduction are presented. In Sect 3, results of the observation are presented, including spectra and intensity maps of the CO lines, intensity and position measurements of the continuum emission and identification of the detected molecular lines. In Sect 4, we discuss the properties of the $^{12}$CO $J=6$$-$$5$ line. Finally the results of the present observation are summarized in Sect 5.


\section{Details of Observations and Data Reduction}
Interferometric observations of CRL~618 were made with SMA on February 18, 2005. Data were obtained under very good atmospheric conditions with a zenith optical depth at 230 GHz of 0.05 on average. The instrument has been described in detail by \citet{ho04} and \citet{oha05}. With the capability of SMA enabling us to use two different receivers at a time, we observed simultaneously in four different frequency bands: 690.572--692.549 GHz, 680.572--682.550 GHz, 219.598--221.575 GHz and 229.600--231.576 GHz. These frequency ranges cover three CO lines: the $^{12}$CO $J=6$$-$$5$ , $^{12}$CO $J=2$$-$$1$  and $^{13}$CO $J=2$$-$$1$ lines. The array used 6 elements (with a diameter of 6m) in the compact array configuration. The baseline length ranged from 10 to 70m. The field of view of a single antenna was 18$''$ and 54$''$ at 690 and 230 GHz, respectively. The observation was interleaved every 15 minutes with nearby gain calibrators, 3c111 and Titan, to track the phase variations over time. Because the angular size of Titan was roughly 0.88$''$$\times$0.68$''$, Titan was not an ideal point source compared with the small size of the beam ($\sim1''$--3$''$) of the present observations. Therefore, in the gain calibration, we used the ``non-point'' flag of the ``gain-cal'' command in the IDL--MIR package, which enabled us to use a planet/satellite as a calibrator (IDL--MIR is a data reduction software developed by the SMA project). The non-point flag function assumes the calibrator is a uniform disk; the size and flux of the calibrator are automatically retrieved and calculated by the internal theoretical calculation. The absolute flux calibration was determined from observations of Ganymede, and was approximately accurate to within 15\%. We adopted the antenna-based solution in the pass band calibration of the 690 GHz data. With enough signal to noise ratio on each channel, the antenna-based bandpass solution is robust and consistent with baseline-based calibration results. For the 230 GHz data, we adopted a standard calibration method using the baseline-based solution. The final map has an accumulated on-source observing time of about 4 hours. The single-sideband system temperature ranged from 2000 to 3000 K at 690 GHz and from 150 to 200 K at 230 GHz, depending on the atmospheric conditions and the telescope direction. The SMA correlator had a bandwidth of 2 GHz with a resolution of 0.812 MHz. The velocity resolution was 0.35 km~s$^{-1}$ at 690 GHz and 1.0 km~s$^{-1}$ at 230 GHz. The phase center of the map was taken at R.A.$=04^{\rm h}42^{\rm m}53.67^{\rm s}$, decl.$=36^{\circ}06'53.2''$ (J2000). Image processesing of the data was performed with the MIRIAD software package \citep{sau95}. We used robust weighting for the 690 GHz data. The robust weighting of the visibility data, which is an optimized compromise between natural and uniform weighting, gave a 1.0$''$ $\times$ 0.8$''$ CLEAN beam at 690 GHz with a position angle of $-12^{\circ}$. For the $^{12}$CO and $^{13}$CO $J=2$$-$$1$ lines, a super-uniform weighting of visibility data (with a suppression region of $8''\times8''$) was adopted to enhance the angular resolution, and this weighting gave a $2.6''\times2.4''$ CLEAN beam with a position angle of $+11.2^{\circ}$. The width of the bands for the continuum maps at 690 and 230 GHz were roughly 1.5 and 1.9 GHz, respectively.


\section{Results}
\subsection{Spectra of CO lines}
Figure 2 shows spatially integrated spectra of the observed CO lines. The line emission was integrated over a circular region with a diameter of 15$''$. The continuum flux was calculated in the line-free channels, and the continuum emission was subtracted from the spectra to retrieve only the line emission. As reported in previous papers, the $^{12}$CO $J=2$$-$$1$ line clearly shows the two different components \citep[i.e., high velocity pedestal component and low velocity strong component; see, e.g.,][]{buj88,buj01,cer89,ner92,mei98,san04,san04a}. In comparison with a single dish spectrum \citep{buj88}, there is no significant loss of the flux in the high velocity component of the $^{12}$CO $J=2$$-$$1$ line, but roughly 45\% of the flux of the low velocity component is resolved out. In fact, the line profile of the high velocity component of the $^{12}$CO $J=2$$-$$1$ line is quite similar with that obtained in previous single dish observations, but the peak intensity of the low velocity component is clearly lower than values of the single dish measurements.

The $^{12}$CO $J=6$$-$$5$ and $^{13}$CO $J=2$$-$$1$ lines have been interferometrically observed for the first time, and, at a glance, these lines exhibit different line profiles compared with the $^{12}$CO $J=2$$-$$1$ line. The high velocity component of the $^{12}$CO $J=6$$-$$5$ line is clearly detected, exhibiting the maximum line width of about 350 km~s$^{-1}$ at the 0 intensity level, while the low velocity component is relatively weak. This weak intensity seems to be caused by missing flux resolved out by the interferometry. In fact, the peak intensity of the $^{12}$CO $J=$6--5 line measured by a single dish observation is 320 Jy \citep{her02}, while the peak intensity of the $^{12}$CO $J=$6--5 line measured in the present observation is 55.3 Jy. In comparison with the Herpin's single dish observation, roughly 80\% of the flux of the low velocity component of the $^{12}$CO $J=$6--5 line is resolved out in the present observation, even though there is no significant missing flux in the high velocity component of the $^{12}$CO $J=6$$-$$5$ line. The large amount of missing flux of the low velocity component implies that the emission region of the low velocity component of the $^{12}$CO $J=6$$-$$5$ line is largely extended. 

The $^{13}$CO $J=2$$-$$1$ line exhibits only the low velocity component, presumably related to the sensitivity. In comparison with a single dish observation \citep{buj88}, there is no significant flux loss in the low velocity component of the $^{13}$CO $J=2$$-$$1$ line. In all three spectra of the CO lines, we can see an absorption feature at $V_{\rm lsr}\sim-40$ km~s$^{-1}$. Although it is somewhat difficult to see the absorption feature in the $^{13}$CO $J=2$$-$$1$ line (at least, there is no absorption below the continuum level), the line profile of the $^{13}$CO $J=2$$-$$1$ line is certainly asymmetric: it shows a weaker blue side compared with the red one (this is clearly seen in the $p$$-$$v$ diagram given in a later section; see, Figure 8). Therefore there is in fact absorption of the blue side of the low velocity component in the $^{13}$CO $J=2$$-$$1$ line. The profile of the low velocity component showing a spiky feature is somewhat different from that of the $^{12}$CO $J=2$$-$$1$ line showing a parabolic profile. This difference is presumably due to a relatively thin optical depth of the $^{13}$CO $J=2$$-$$1$ line. The spiky profile possibly suggests a complex kinematics in the inner part the molecular envelope. In the bottom panel of Figure 2, we can see a broad emission feature at $V_{\rm lsr}\sim-300$ km~s$^{-1}$, which can be explaned by leakage of the strong $^{12}$CO $J=2$$-$$1$ line lying in the other side band.  This conclusion is based on the fact that he peak velocity of this feature exactly corresponds to the backside of the $^{12}$CO $J=2$$-$$1$ line, the line profile is very similar with that of the $^{12}$CO $J=2$$-$$1$ line, and also because no possible lines are found in molecular line catalogs. (Note that this leakage problem has already been fixed, and the current SMA system has no problems even in the case of very strong lines such as CO lines.)

\subsection{Intensity maps of CO lines}
Figure 3 shows the velocity channel maps of the $^{12}$CO $J=6$$-$$5$ line. The background gray scale represents the continuum emission at 690 GHz, and the origin of the map is taken at the intensity peak of the continuum emission. In Figure 3, we can clearly see a velocity gradient in the E--W direction as well as the $^{12}$CO $J=2$$-$$1$ map given by \citet{san04}. Interestingly, we cannot see the cavity structure found by \cite{san04} in the $^{12}$CO $J=2$$-$$1$ line, although the angular resolution of the present observation in the $^{12}$CO $J=6$$-$$5$ line roughly equals that of S\'anchez Contreras's observation in the $^{12}$CO $J=2$$-$$1$ line. Both ends of the high velocity component (i.e., from $-166$ to  $-116$ km~s$^{-1}$ and from $35$ to $115$ km~s$^{-1}$) are clearly extended in the E--W direction up to roughly $\pm2.5''$ from the nebula center. This size of the extension in the E--W direction is the same as that seen in the $^{12}$CO $J=2$$-$$1$ line. 

In Figure 4, we present the velocity integrated intensity maps of the $^{12}$CO $J=6$$-$$5$ line superimposed on the HST WFPC2 H$_{\alpha}$$+$continuum image \citep{tra02}. The cross indicates the intensity peak of the radio continuum emission at 690 GHz. The direction of the elongation of the $^{12}$CO $J=6$$-$$5$ feature coincides with that of the optical feature seen in the HST image. The emission region of the high velocity component of the $^{12}$CO $J=6$$-$$5$ line is limited to the optically faint region (i.e. the vicinity of the nebula center). In comparison with the maps in \citet{san04}, the spatial size of the emission region of the low velocity component of the $^{12}$CO $J=6$$-$$5$ line is smaller than that of the $^{12}$CO $J=2$$-$$1$ line. This small size is due presumably to the large missing flux. 

Figure 5 shows the channel velocity maps of the $^{13}$CO $J=2$$-$$1$ line. In contrast to the $^{12}$CO $J=6$$-$$5$ line, the velocity gradient is not clear. This is due to a low signal-to-noise ratio in the high velocity component of the $^{13}$CO $J=2$$-$$1$ line. In Figures 6 and 7, we present the velocity channel maps of the $^{12}$CO $J=2$$-$$1$ line. We present the maps of the low velocity component in Figure 6, and also present those of the high velocity component in Figure 7. The features of the $^{12}$CO $J=2$$-$$1$ line seen in Figures 6 and 7 are in good agreement with the maps given by \citet{san04}; we can clearly see a velocity gradient in the E--W direction. A cavity structure seen in S\'anchez Contreras's maps is not clear in Figure 6. This is due to the low angular resolution of the present observation. In Figure 7, both ends of the high velocity component are extended to in the E--W direction up to $\pm2.5''$ from the nebular center. This is also consistent with the results of \citet{san04}. We cannot see the spherical halo seen in the S\'anchez Contreras's maps due to a lack of sensitivity. In fact, a 1 $\sigma$ level in Figure 2b in \citet{san04} is 45 mJy beam$^{-1}$ (beam size is 1.1$'' \times$0.9$''$); in contrast, a 1 $\sigma$ level in Figure 6 is 100 mJy beam$^{-1}$ (beam size is 3.3$'' \times$2.8$''$). Missing flux, in principle, could be a reason for the absence, but we recovered 55\% of the flux of the low velocity component, and this recovery rate is larger than the 40\% of Sanchez Contreras's observation. 

For a better understanding of kinematical properties, we present position-velocity ($p$$-$$v$) diagrams in Figure 8. The present data are superimposed on the $p$$-$$v$ diagram of the $^{12}$CO $J=2$$-$$1$ line taken from \citet{san04}. The left and right columns represent $p$$-$$v$ diagrams in the axial and perpendicular directions, respectively. In the middle two panels, we can see that the emission of the $^{12}$CO $J=2$$-$$1$ line is not detected in the high velocity ends due to a lack of sensitivity. The feature of the $^{12}$CO $J=6$$-$$5$ line is basically consistent with that of the $^{12}$CO $J=2$$-$$1$ line, even though the spatial size of the low velocity component of the $^{12}$CO $J=6$$-$$5$ line is smaller than that of the $^{12}$CO $J=2$$-$$1$ line.

\subsection{Continuum emission}
The continuum emission of CRL~618 was detected both at 230 and 690 GHz. The continuum emission at 690 GHz has been interferometrically observed for the first time. These continuum emissions seem to exhibit a point-like feature in the intensity maps, but the $uv$--distance versus amplitude plot of the 690 GHz continuum emission shows that the amplitude increases with decreasing $uv$--distance (although we do not present the plot here), suggesting the 690 GHz continuum flux is partially resolved out. The integrated fluxes of the continuum emission are 2.0 and 2.7 Jy at 230 and 690 GHz, respectively. The 230 GHz continuum flux is consistent with the latest measurement by \citet{san04}. The spectral index calculated by the 230 and 690 GHz measurements is $\nu^{0.3}$. This is not consistent with the spectral index expected from optically thin free-free emission  ($\nu^{-0.1}$). Incidentally, the turning-off point of the spectrum of the central HI\hspace{-.1em}I region is roughly at 45~GHz \citep{kna93}. Because the angular size of the central ionized region is 0.2$''\times$0.4$''$, the free-free emission is not expected to be resolved by our synthesized beam. Therefore, the extended feature in the 690 GHz continuum emission presumably originates in the dust component in the envelope. The positions of the intensity peak of the continuum emission measured by two-dimensional Gaussian function fitting are 4$^{\rm h}$42$^{\rm m}$53.58$^{\rm s}$, 36$^{\circ}$06$'$53.4$''$ (J2000) at 230 GHz (FWHM and position angle are 3.27$''$$\times$2.84$''$ and $-19.2^{\circ}$, respectively), and 4$^{\rm h}$42$^{\rm m}$53.59$^{\rm s}$, 36$^{\circ}$06$'$53.3$''$ (J2000) at 690 GHz (FWHM and position angle are 1.07$''$$\times$0.88$''$ and of $-13.0^{\circ}$, respectively). These positions coincide well with that determined by a low frequency observation \citep{wyn77}, but are slightly shifted from the intensity peak of the CO $J=6$$-$5 emission (see lower panel in Figure 4. We will discuss this matter later in Sect 4).

\subsection{Other detected lines}
In Figure 9 we present a spectrum including all lines detected in the 230 GHz bands. The continuum emission was subtracted in Figure 9. In total we have detected 36 emission lines above the 5 $\sigma$ level, and 24 out of the 36 detected lines are identified to known molecular lines. We used ``WinSpectra'', the line identification software developed by the ODIN group at the University of Calgary to identify the lines. WinSpectra uses the JPL and CDMS catalogs \citep{mul01,mul05,pea05} in a built-in form to retrieve the molecular data. The results of the line identification are summarized in Table 1. Most lines seen in Figure 9 exhibit a P Cygni profile as reported in previous observations \citep{buj88,wyr03,par04,par07}. This suggests that most of the detected lines originate from an expanding and accelerating molecular envelope \citep[see, e.g.,][]{wyr03}. The frequencies in Table 1 represent the rest frequencies (for identified lines) or laboratory frame frequencies, corrected using $V_{\rm lsr}=-23$ km~s$^{-1}$ (for unidentified lines). ``U'' in Table 1 indicates unidentified lines. The radial velocities of the identified lines in Table 1 are velocities at the intensity peak. No recombination lines lye within the observed frequency ranges. We inspected intensity maps of all detected molecular lines, but we did not find any extended features.


\section{Discussion}
In this section, we focus on the properties of the $^{12}$CO $J=6$$-$$5$ line. In particular, we discuss the structure probed by the $^{12}$CO $J=6$$-$$5$ line, and the large missing flux of the low velocity component of the $^{12}$CO $J=6$$-$$5$ line. 

A notable finding in the present observation is that the cavity structure, which has been found by the S\'anchez Contreras's observation in the $^{12}$CO $J=2$$-$$1$ line, is not clear in the $^{12}$CO $J=6$$-$$5$ maps, although the beam sizes are almost the same with the S\'anchez Contreras's observation. An immediate interpretation for the missing cavity is that the emission of the $^{12}$CO $J=6$$-$$5$ line corresponding to the cavity wall is resolved out by the interferometry. However, we have to be careful in this interpretation, because the cavity wall probed by the $^{12}$CO $J=2$$-$$1$ line has exhibited thin structure (thickness$\sim1''$). 

In a careful second look of Figure 3, we realize that features seen in velocity-channel maps of the $^{12}$CO $J=6$$-$$5$ seem to trace the south wall of the cavity even though it is not very clear. In Figure 3, for example, the map at $-26$ km~s$^{-1}$ shows an extension to the east-southeast direction from the nebula center, and the map at $-16$ km~s$^{-1}$ shows an extension to the west-southwest direction from the nebula center. In addition, in the lower panel of Figure 4, the intensity peak of the CO emission seems to be shifted ($\sim 0.8''$ to south) from the position of the continuum sources. These phenomenon, presumably, suggest that we detected only the south wall of the cavity in the $^{12}$CO $J=6$$-$$5$ line, and that we did not detect the central dense core and the north wall of the cavity.

However, we still have a couple of questions: (1) what is the source of the missing flux of the $^{12}$CO $J=6$$-$$5$ line, and (2) why is only the north wall detectable in the $^{12}$CO $J=6$$-$$5$ line? In terms of the expanding velocity and the angular size of the structure, only the spherical halo and the slow axial component could be a possible source of the missing flux. The emission from the spherical halo is expected to be partially resolved out if it is detectable in the $^{12}$CO $J=6$$-$$5$ line, because the spherical halo is largely extended (at least, in the $^{12}$CO $J=2$$-$$1$ line). However, an important point is that we cannot expect the spherical halo to be highly excited enough to emit the $^{12}$CO $J=6$$-$$5$ line. The CO lines emitted from a circumstellar spherical halo are usually radiatively excited by infrared emission from the central star and dust component in the envelope (the energy input to the dust component originally comes from the central star. Therefore, we cannot expect a strong IR emission in the outer part of the envelope even if we assume the IR emission from the dust). Therefore, only the cavity wall (slow axial component) may be the possible source of the missing flux, at least, among the known components in the CRL~618 molecular envelope. 

In the current scenario \citep{lee03,san04}, the cavity wall is formed by an interaction between ambient material and the energetic high velocity jet. The shock caused by the energetic jet would be a possible reason for a high temperature/density which would excite the $^{12}$CO $J=6$$-$$5$ line. In fact, the S\'anchez Contreras's model, which has reasonably reproduced their observations in the CO and HC$_3$N lines, has applied 200 K as the temperature of the cavity wall. This temperature is much higher than the excitation temperature of the $^{12}$CO $J=6$$-$$5$ line ($\sim 100$K). 

In our opinion, the following two reasons cause the absence of the north wall of the cavity. First, the temperature and/or density of the north wall is smaller than those of the south wall. Consequently, the north wall exhibits a weaker intensity in the $^{12}$CO $J=6$$-$$5$ line. In fact, the north wall shows a weaker intensity in the $^{12}$CO $J=2$$-$$1$ line \citep{san04}. Second, the different $uv$--coverage between S\'anchez Contreras's and the present observations possibly causes the missing northern cavity wall. The rate of the missing flux strongly depends on the pattern of the $uv$--coverage. In some cases an elongated structure could be resolved out even if it shows thin structure. Presumably, the missing northern cavity is a combination of these two effects. The difference in the temperature/density between the north and south walls, if it is real, leads to the fact that the physical conditions in the CRL~618 molecular envelope is not exactly axial symmetric. The southern part of the envelope (especially, in the innermost part) may exhibit a relatively higher temperature/density as compared with the northern part. This non-axial symmetric distribution of the temperature/density might be caused by the central star (for example, binarity and/or time-variation of post-AGB wind, which are very poorly understood).


\section{Summary}
This paper has reported the results of an SMA observation of the proto-planetary nebula CRL 618 in the $^{12}$CO $J=6$$-$$5$ line. We also observed simultaneously in the $^{12}$CO $J=2$$-$$1$ and $^{13}$CO $J=2$$-$$1$ lines. In addition to the CO lines, we have detected a number of emission lines at 230 GHz. The main results are as follows:

\begin{enumerate}
\item Flux of the high velocity component of the $^{12}$CO $J=6$$-$$5$ line is almost fully recovered, while we lost roughly 80\% of the low velocity component flux. This suggests that the emission region of the low velocity component of the $^{12}$CO $J=6$$-$$5$ line is largely extended.

\item The existence of the cavity structure, which has been found in a previous observation in the $^{12}$CO $J=2$$-$$1$ line, is not clear in the $^{12}$CO $J=6$$-$$5$ maps. Only the south wall of the cavity seems to be detected in the $^{12}$CO $J=6$$-$$5$ line. This result suggests that the physical condition of the molecular envelope of CRL~618 is not exactly axial symmetric.

\item Continuum emission is detected both at 230 and 690 GHz. The flux of the 690~GHz continuum emission seems to be partially resolved out by the interferometry. This possibly suggests that dust emission significantly contaminates the 690~GHz continuum flux.
\end{enumerate}


\acknowledgments
We are grateful to Rob Christensen, Alison Peck and the SMA team for making the 690 GHz campaign possible. This research has been supported by the Academia Sinica Institute of Astronomy \& Astrophysics and the Smithsonian Institute, and has made use of the SIMBAD and ADS databases. The authors thank Jinhua He, Holger M\"ueller and Shuro Takano for their help in the identification of the molecular lines. JN thanks Paul Ho for his constant encouragement.



\begin{figure}
\epsscale{.70}
\plotone{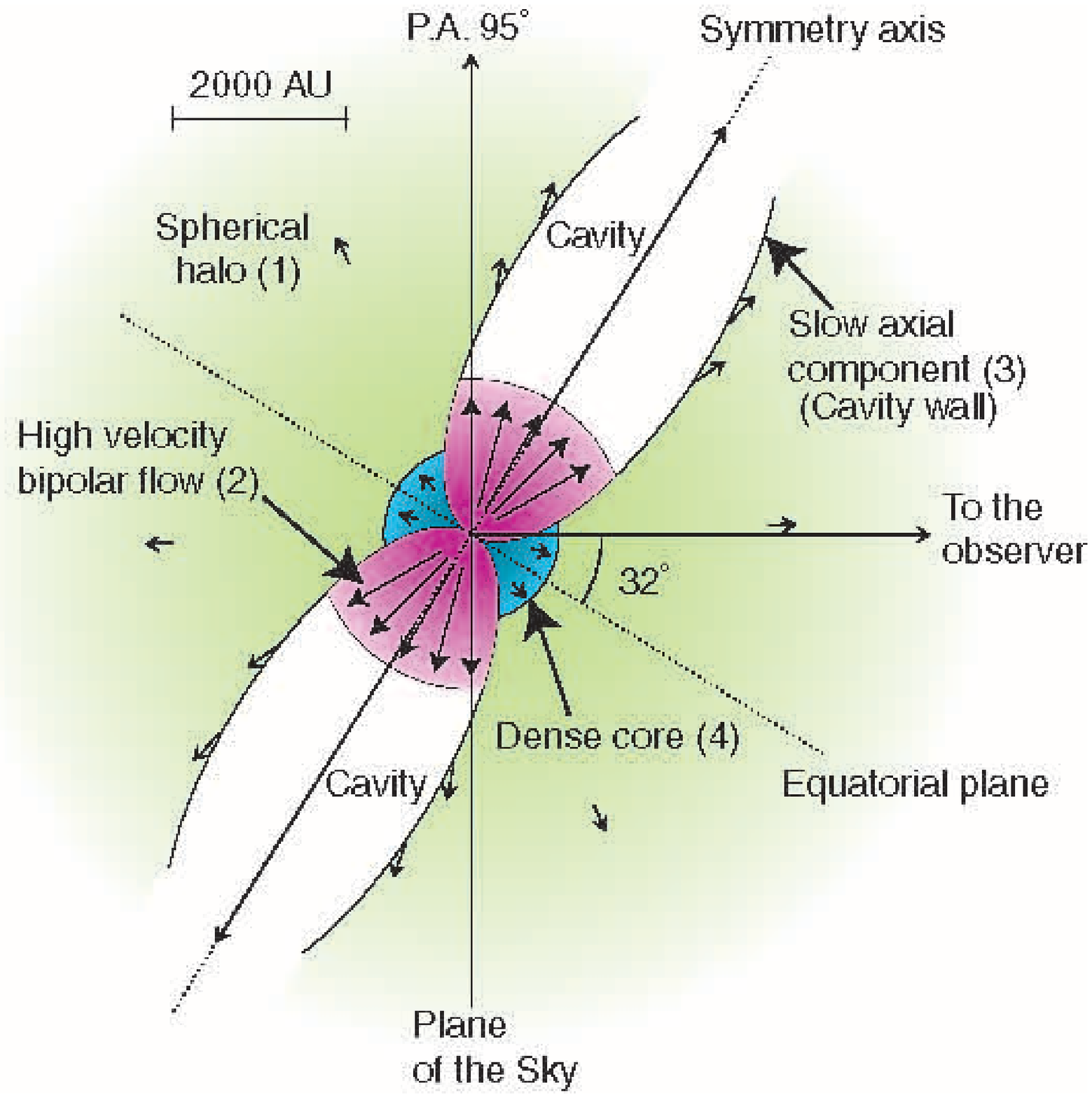}
\figcaption{Schematic view of the molecular envelope of CRL 618 made using examples from Figure 5 in \citet{san04}. The small arrows indicate the velocity field of the envelope. The numbers in parentheses correspond to the numbers of the components referred to in the text (see Section 1). \label{fig1}}
\end{figure}
\clearpage

\begin{figure}
\epsscale{.60}
\plotone{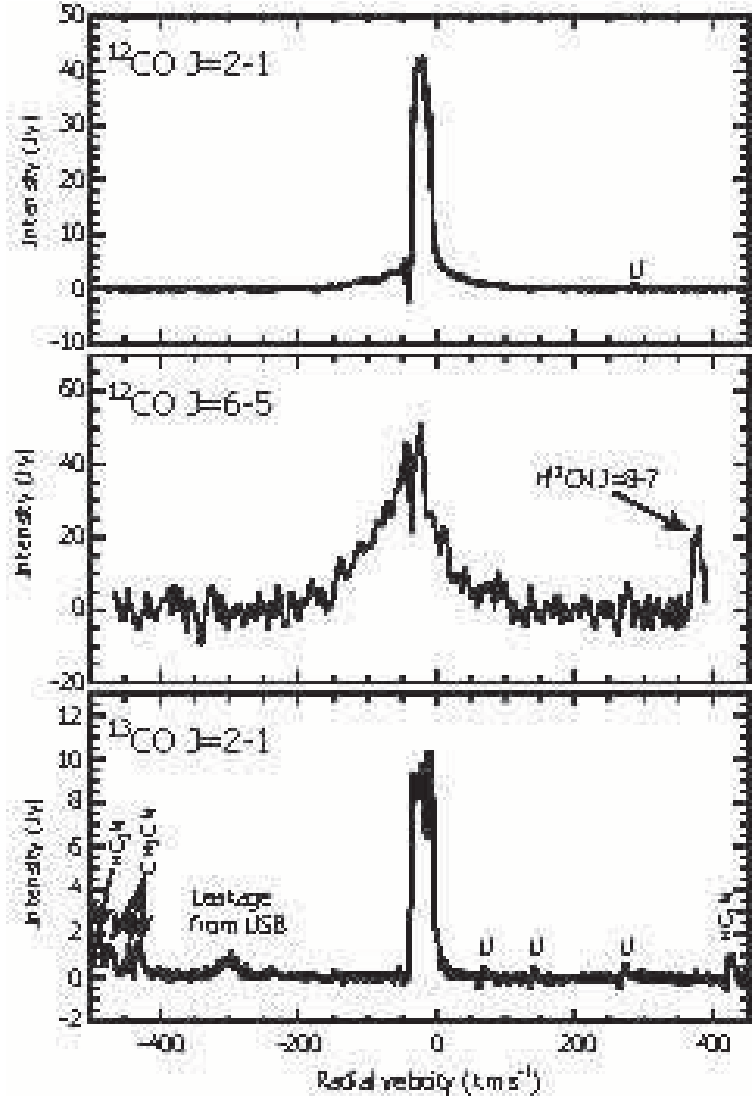}
\figcaption{Spatially integrated spectra of the $^{12}$CO $J=2$--1, $^{12}$CO $J=6$--5 and $^{13}$CO $J=2$--1 lines. The integrated area is a circle with an angular diameter of 15$''$. For the $^{12}$CO $J=6$--5 line, the spectrum is smoothed over every 5 km s$^{-1}$. The other detected lines seen in the spectra are indicated by the name of the molecular species or ``U'' meaning unidentified line. \label{fig2}}
\end{figure}
\clearpage

\begin{figure}
\epsscale{.80}
\plotone{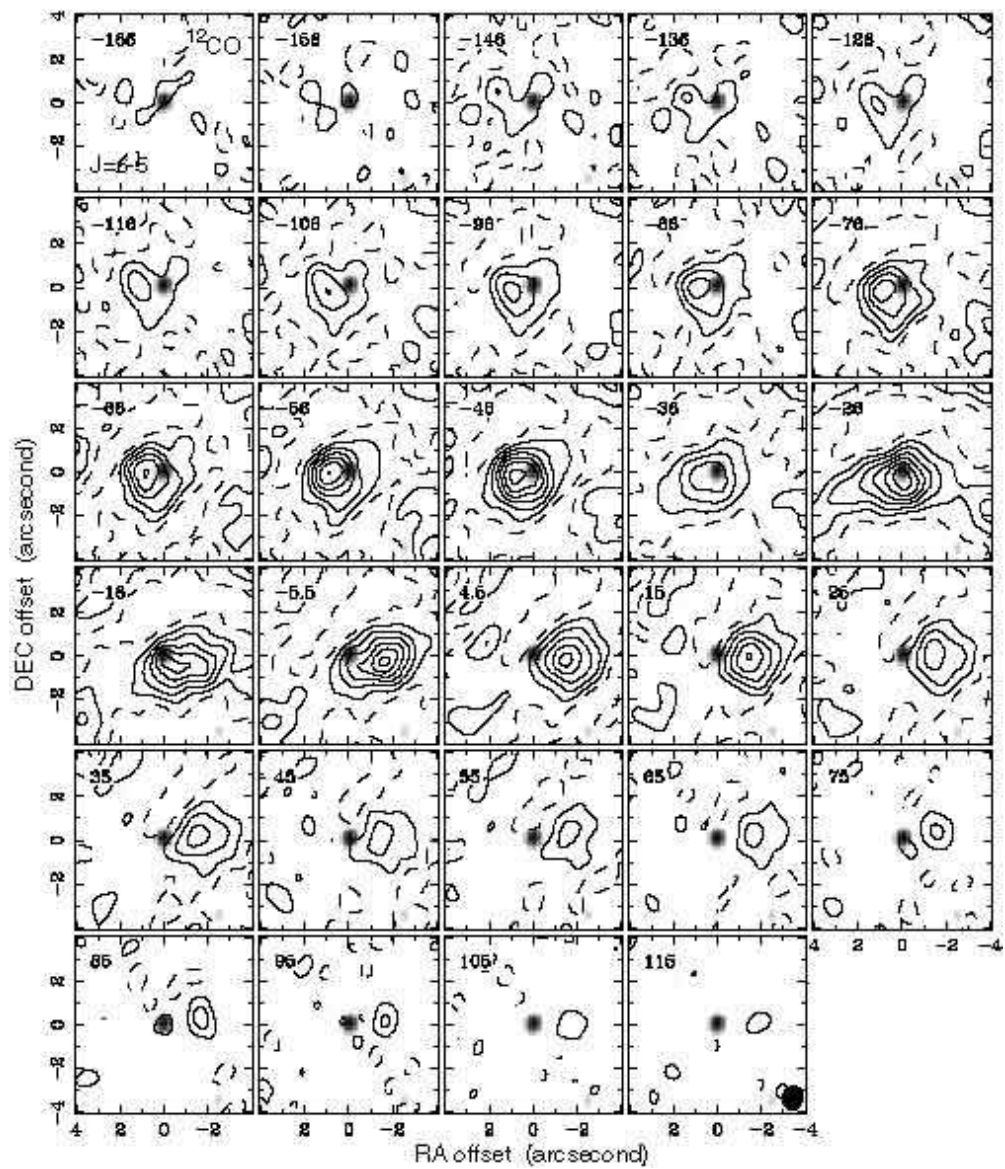}
\figcaption{Velocity channel maps of the CO $J=6$$-$$5$ line. The velocity channels are averaged over 10~km~s$^{-1}$ intervals. The contours start from a 3~$\sigma$ level, and the levels are spaced every 5~$\sigma$. The 1~$\sigma$ level corresponds to $4.42\times10^{-1}$ Jy~beam$^{-1}$. The dashed contour corresponds to $-3~\sigma$. The peak intensity corresponds to 28~$\sigma$. The synthesized beam is indicated in the lower right panel. The background gray scale represents the intensity of the 690~GHz continuum emission. \label{fig3}}
\end{figure}
\clearpage

\begin{figure}
\epsscale{.60}
\plotone{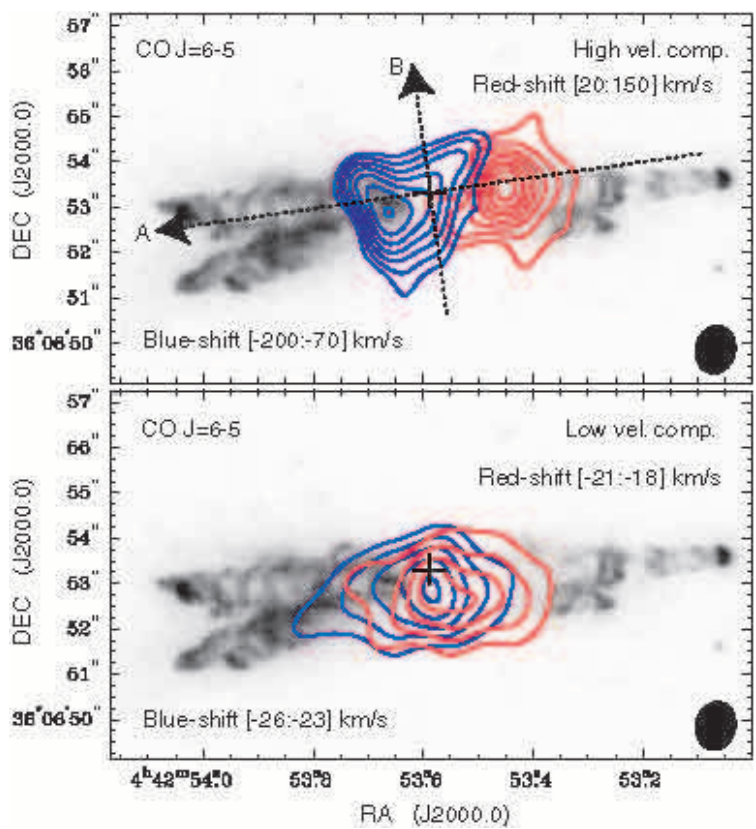}
\figcaption{Velocity integrated intensity maps of the CO $J=6$$-$5 line superimposed on the {\it HST} WFPC2 H$_{\alpha}+$continuum image \citep{tra02}. The red- and blue-contours represent the red- and blue-shifted parts of the high and low velocity components. The velocity ranges of the integration are given in the panels. The lowest contour corresponds to a 3~$\sigma$ level, and the levels are spaced every 4~$\sigma$; a 1~$\sigma$ level for the upper and lower panels are $1.2\times10^{-1}$ Jy~beam$^{-1}$ and $6.9\times10^{-1}$ Jy~beam$^{-1}$, respectively. The synthesized beam size is given in the lower-right corners. The dotted arrows, A and B, indicate the directions along which the $p$$-$$v$ cuts shown in Figure 8 were taken. \label{fig4}}
\end{figure}
\clearpage

\begin{figure}
\epsscale{.80}
\plotone{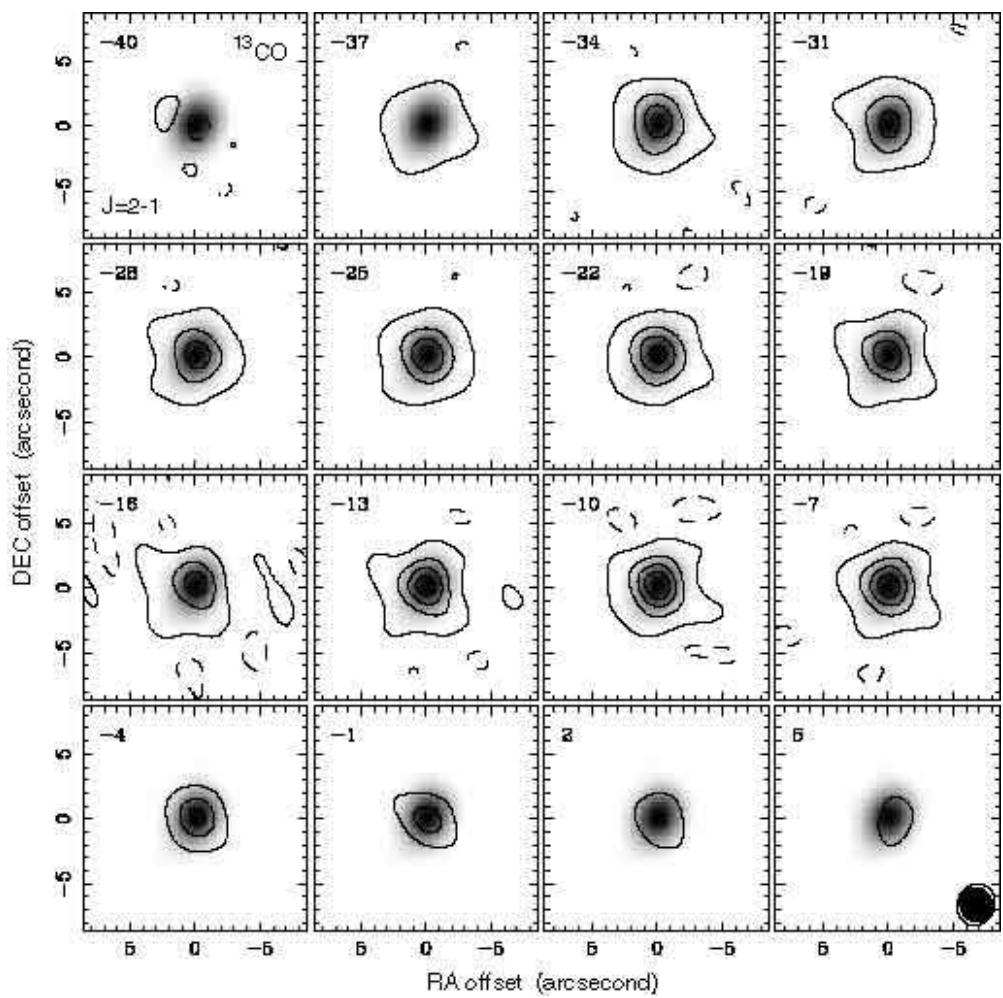}
\figcaption{Velocity channel maps of the $^{13}$CO $J=2$$-$$1$ line. The velocity channels are averaged over 3~km~s$^{-1}$ intervals. The contours start from a 3~$\sigma$ level, and the levels are spaced every 10~$\sigma$. The 1~$\sigma$ level corresponds to $9.5\times10^{-2}$ Jy~beam$^{-1}$. The dashed contour corresponds to $-3~\sigma$. The intensity peak corresponds to 40~$\sigma$. The background gray scale represents the intensity of the 1~mm continuum emission. The synthesized beam is indicated in the lower right panel; the outer unfilled and inner filled ellipses represent the beam sizes for the line and continuum maps, respectively. \label{fig5}}
\end{figure}
\clearpage

\begin{figure}
\epsscale{.80}
\plotone{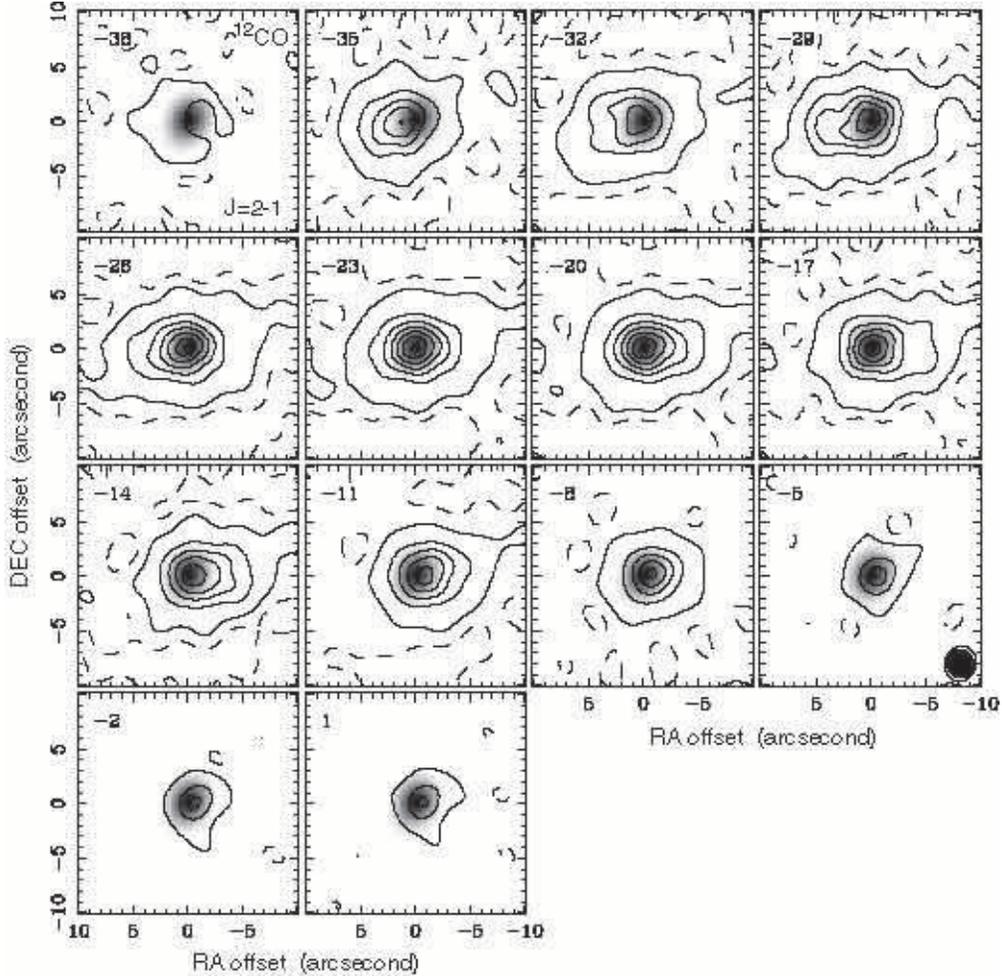}
\figcaption{Velocity channel maps of the low velocity component of the CO $J=2$$-$$1$ line. The channel velocities are given in the upper left corners of each panel. The velocity channels were averaged over 1~km~s$^{-1}$ intervals. The contours start from a 3~$\sigma$ level, and the levels are spaced every 20~$\sigma$. The dashed contour corresponds to $-3~\sigma$. The 1~$\sigma$ level corresponds to $1.03\times10^{-1}$ Jy~beam$^{-1}$. The peak intensity corresponds to 120~$\sigma$. The background gray scale represents the intensity of the 1~mm continuum emission. The synthesized beam is indicated in the lower right panel; the outer unfilled and inner filled ellipses represent the beam sizes for the line and continuum maps, respectively. \label{fig6}}
\end{figure}
\clearpage

\begin{figure}
\epsscale{.80}
\plotone{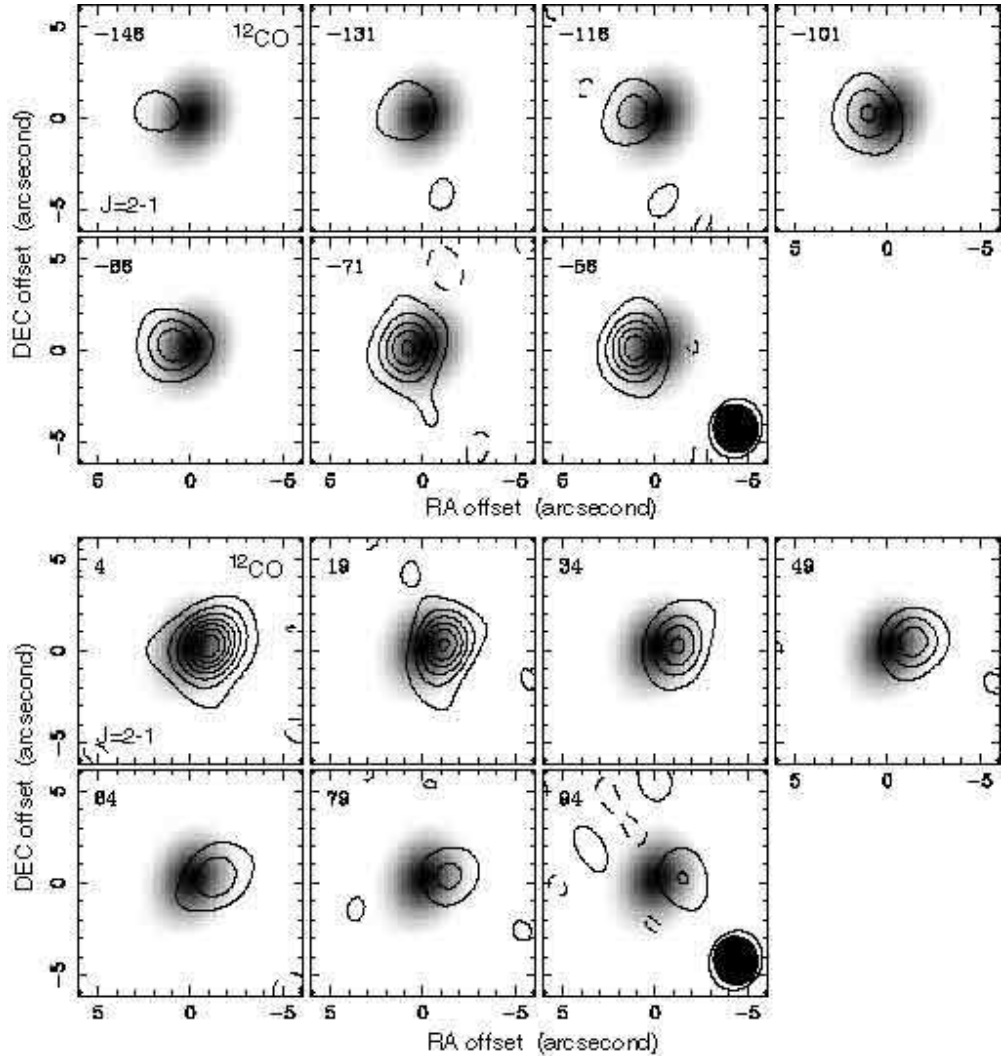}
\figcaption{Velocity channel maps of the high velocity component of the CO $J=2$$-$$1$ line. The velocity channels are averaged over 1~km~s$^{-1}$ intervals. The contours start from a 3~$\sigma$ level, and the levels are spaced every 10~$\sigma$. The dashed contour corresponds to $-3~\sigma$. The 1~$\sigma$ level, background gray scale and synthesized beam are the same as in Figure 6. \label{fig7}}
\end{figure}
\clearpage

\begin{figure}
\epsscale{.80}
\plotone{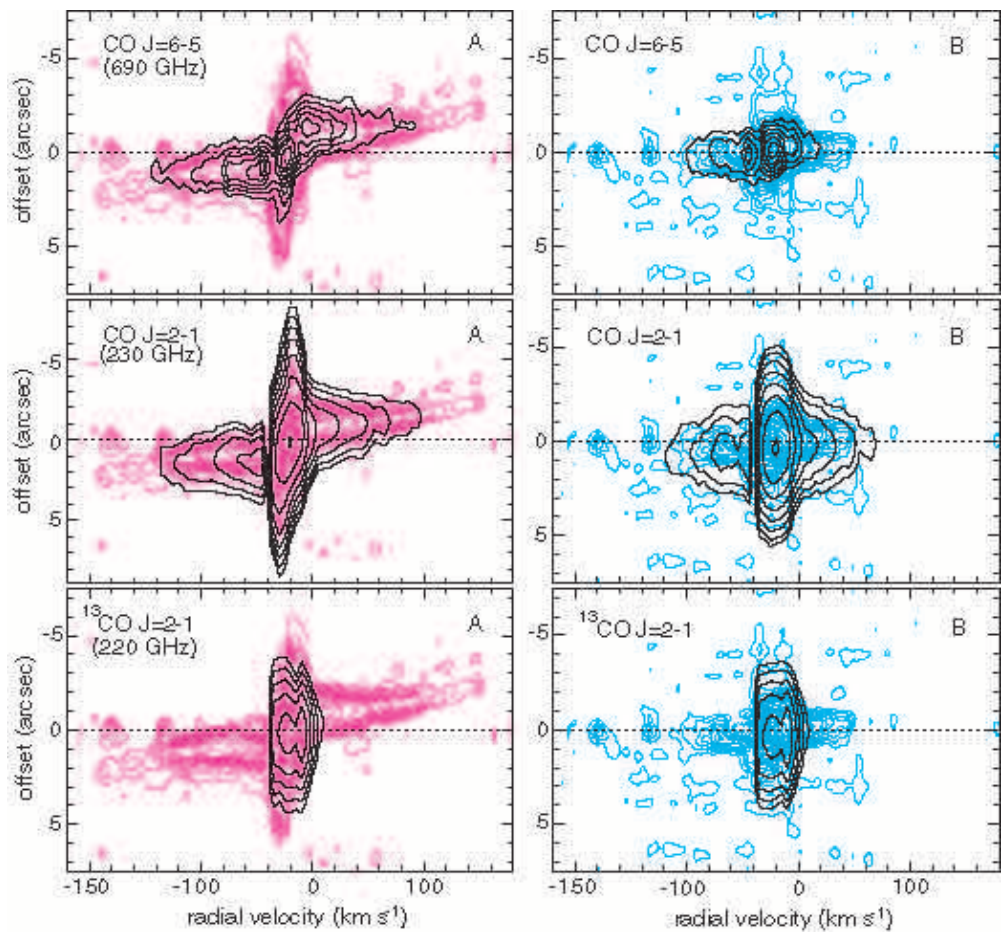}
\figcaption{Position-velocity diagrams of the CO $J=6$--5, CO $J=2$--1 and $^{13}$CO $J=2$--1 lines superimposed on the similar diagram of the $^{12}$CO $J=2$--1 line taken from \citet{san04}. ``A'' and ``B'' mean the direction of cut indicated in Figure 4. The contours start at a 3~$\sigma$ level. A linear scale is used for the top panel, and the increment of the contours is every 1.75 Jy~beam$^{-1}$. A logarithmic scale is used for the middle and bottom panels. The dashed horizontal lines represent the origin of the offset axes. \label{fig8}}
\end{figure}
\clearpage

\begin{figure}
\epsscale{.80}
\plotone{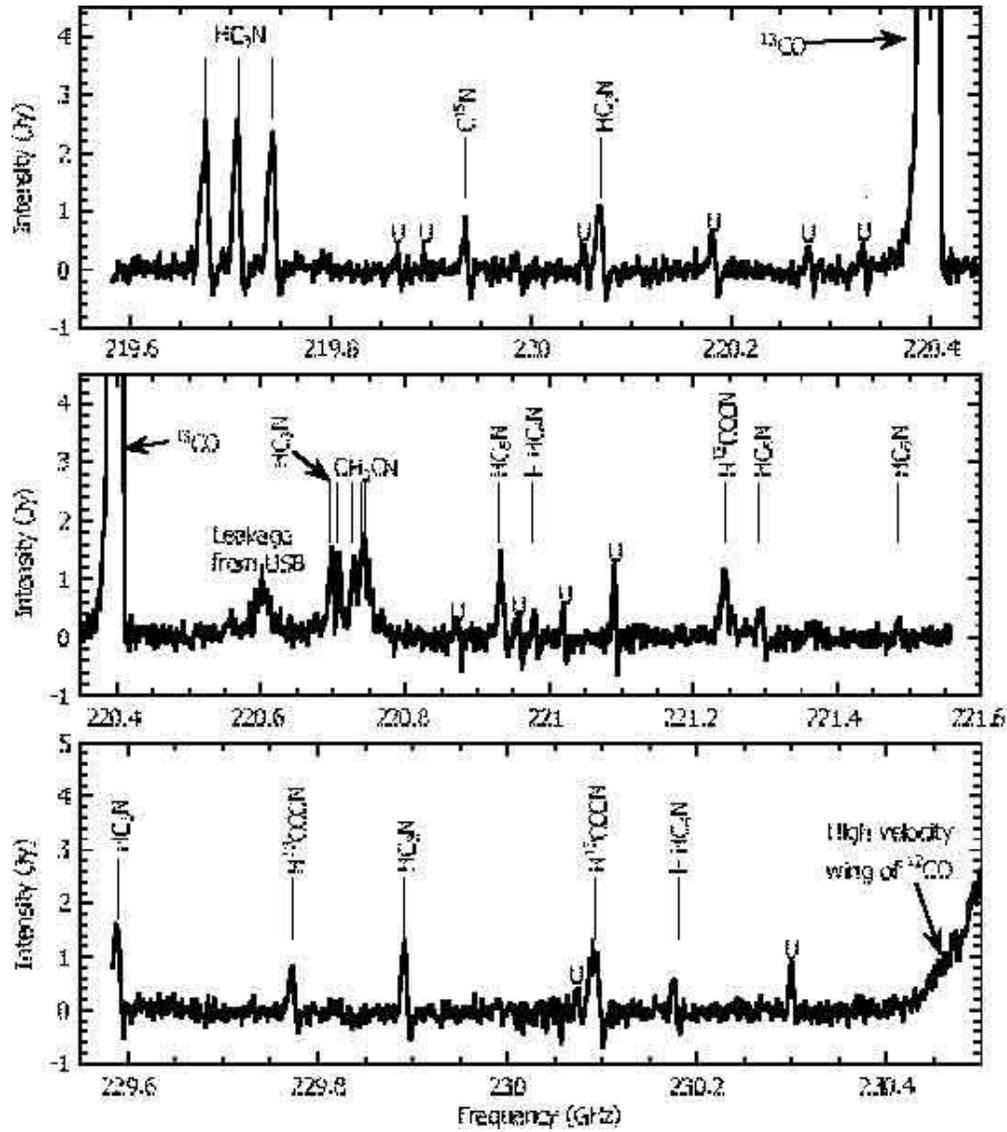}
\figcaption{Spatially integrated spectra in the 1~mm band. The integrated area is the same as Figure 1. The horizontal axis represents the laboratory frame frequency, corrected using $V_{\rm lsr}=-23$ km s$^{-1}$. The vertical solid bars indicate the rest frequencies of identified lines. ``U'' means unidentified line. \label{fig9}}
\end{figure}
\clearpage


\begin{deluxetable}{lllll}
\tablecolumns{5}
\tablewidth{0pc}
\tablecaption{Line Detections}
\tablehead{
\colhead{Frequency}    &  \colhead{Species} &   \colhead{Transition}  &
\colhead{$V_{\rm lsr}$} & \colhead{$S$} \\
\colhead{(MHz)} & \colhead{}   & \colhead{}    & \colhead{(km s$^{-1}$)} &
\colhead{({\tiny Jy km s$^{-1}$}})}
\startdata
219675.114 & HC$_3$N & $J=24$--23 $\nu_7 =2$ $l=0$ & $-22.0$ & 26.3 \\
219707.349 & HC$_3$N & $J=24$--23 $\nu_7 =2$ $l=2$e & $-22.1$ & 25.4 \\
219741.866 & HC$_3$N & $J=24$--23 $\nu_7 =2$ $l=2f$& $-23.1$ & 25.1 \\
219866.826 & U &  & --- & 1.37 \\
219892.538 & U &  & --- & 1.13 \\
219934.820 & C$^{15}$N & $N=$2--1 $J=5/2$--$3/2$ & $-22.4$ & 4.89  \\
220050.478 & U &   & --- & 2.89  \\
220070.185 & HC$_3$N & $J=24$--23 $\nu_7 =3$ $l=1e$  & $-20.2$  &  10.1 \\
220181.972 & U &   & --- &  4.87 \\
220278.206 & U &   & --- & 2.54 \\
220332.566 & U &   & --- & 2.54 \\
220398.681 & $^{13}$CO & $J=2$--1  & $-10.0$ & 310.0 \\
220699.688 & HC$_3$N & $J=24$--23 $\nu_7 =3$ $l=1f$ & $-21.2$ & 17.4 \\
220709.024 & CH$_3$CN & $J_{\rm K}=12_{3}$--11$_3$  & $-22.5$ & 14.9 \\
220730.266 & CH$_3$CN & $J_{\rm K}=12_{2}$--11$_2$  & $-21.6$ & 16.5 \\
220743.015 & CH$_3$CN & $J_{\rm K}=12_{1}$--11$_1$  & $-22.3$ & --- \\
220747.265 & CH$_3$CN & $J_{\rm K}=12_{0}$--11$_0$  & $-25.5$ & --- \\
220868.094 & U &  & --- & 2.60 \\
220932.401 & HC$_5$N & $J=83$--82 $\nu =0$  & $-23.5$ & 15.8 \\
220957.716 & U &   & --- & 3.28 \\
220979.090 & $l$--HC$_4$N & $J=48$--47 & $-22.9$ & 3.08 \\
221020.158 & U &   & --- & 3.34 \\
221090.680 & U &   & --- & 8.77 \\
221246.240 & H$^{13}$CCCN & $J=25$--24 $\nu_7 =1$ $l=1e$ & $-18.3$  & 19.8 \\
221295.423 & HC$_5$N & $J=83$--82 $\nu_{11}=1$ $l=1f$ & $-23.3$ & 5.15 \\
221486.033 & HC$_5$N & $J=83$--82 $\nu_{11}=1$ $l=1e$ & $-22.8$  & 2.42 \\
229589.088 & HC$_3$N & $J=25$--24 $\nu_7=3$ $l=3f$ & $-19.9$ & 16.5 \\
229772.876 & H$^{13}$CCCN & $J=26$--25 $\nu_7 =1$ $l=1e$ & $-23.7$ & 6.61 \\
229891.277 & HC$_3$N & $J=25$--24 $\nu_7=3$ $l=1e$ & $-23.6$ & 9.46 \\
230074.654 & U &  & --- & 2.19 \\
230093.224 & H$^{13}$CCCN & $J=26$--25 $\nu_7=1$ $l=1f$ & $-18.8$ & 14.6 \\
230180.393 & $l$--HC$_4$N & $J=50$--49 & $-17.4$  & 4.04 \\
230299.796 & U &  & --- & 5.81 \\
230538.000 & CO & $J=2$--1  & $-20.0$ & 1541.4 \\
690552.098 & H$^{13}$CN & $J=8$--7  & $-19.6$ & 346.4 \\
691473.076 & CO & $J=6$--5  & $-22.0$ & 4647.8 \\
\enddata
\end{deluxetable}


\end{document}